\documentclass[twocolumn,preprintnumbers,amsmath,amssymb]{revtex4}

\usepackage{graphicx}
\usepackage{dcolumn}
\usepackage{bm}
\usepackage{epsfig}
\usepackage{amssymb}
\begin{document}

\preprint{}

\title{Response to arXiv:1005.2615}

\author{J.I. 
Collar$^{a}$
}
\author{D.N. McKinsey$^{b}$
}

\address{ 
$^{a}$Enrico 
Fermi Institute, KICP and  Department of Physics, University of Chicago, Chicago, IL 60637\\
}
\address{ 
$^{b}$Department of Physics, Yale University, New Haven, CT 06520\\
}
\maketitle

The XENON100 collaboration has offered a Reply \cite{reply} to our
Comments \cite{v3} on their first results \cite{xenon100}.  We find it
inadequate, for more than one reason.  First, we deem that clear
efforts are made in \cite{reply} to distort what are otherwise several
straightforward points made by us in \cite{v3}.  We believe that these
cannot be missed by an attentive reader.  In order to keep the
discussion brief, rather than disputing these points one by one, we
concentrate below on a number of broader related issues.  Second, we
sense an avoidance of the main criticism contained in our Comments. 
We start by restating this criticism below, and conclude this brief note with a
challenge to the XENON100 collaboration.

\begin{enumerate}

\item It is simply not legitimate to base a dark matter cross-section
limit on presumed Poisson fluctuations in S1 in an energy range where
there are no measurements of $\mathcal{L}_{\text{eff}}$ (i.e. below 4
keV$_{r}$).  If we accept as a community that it is not proper to 
take advantage of the Poisson fluctuations from a completely unquantified
light source, the controversy about the behavior of
$\mathcal{L}_{\text{eff}}$ at low energies will be less relevant.  In
any case, we recall that $\mathcal{L}_{\text{eff}}$ is the ratio
between the (poorly understood) nuclear recoil scintillation yield and
the (well understood and finite) electron recoil scintillation yield
from 122 keV gamma rays.  In the limit of low energy,
$\mathcal{L}_{\text{eff}}$ has to go to zero, because eventually there
is not enough energy for the nuclear recoil to excite the xenon atom
to its first excited state.  The numerator in
$\mathcal{L}_{\text{eff}}$ must go to zero, and the denominator is
finite.  The question is where, not if, $\mathcal{L}_{\text{eff}}$
goes to zero.  While the XENON100 collaboration agrees
with us that more accurate measurements of $\mathcal{L}_{\text{eff}}$
are needed, when setting a limit on a dark matter cross-section it is
most responsible to be conservative about the assumptions going into
the sensitivity.

\item Note that to obtain our limit curves, in an attempt to replicate
the XENON100 claimed limits, we did in fact include S1 fluctuations,
and we succeed in reproducing their limits under the same set of
questionable assumptions.  Section 4 in \cite{reply} is rife with
misleading references to what we trust is rather clearly stated in the
main text and Appendix of our \cite{v3}.  We encourage other
researchers to attempt to replicate the XENON100 limit curve at low
WIMP masses and to assess the effect of uncertainties in
$\mathcal{L}_{\text{eff}}$.  We firmly maintain that Fig.\ 5 in our
Comments is representative of the degree of uncertainty that can be
presently assigned to light-WIMP studies using LXe as a target.

\item The Reply by XENON100 \cite{reply} calls attention to a
systematic correction applied in Manzur {\it et al.} \cite{manzur},
based on trigger threshold.  The work by Manzur {\it et al.} includes
measurements in both single phase (scintillation only) and two-phase
(scintillation plus charge) modes.  Operating in two phase mode, the
trigger was based on proportional scintillation from charge, allowing
a trigger threshold well below the analysis threshold.  Manzur {\it et
al.} operated in both modes in order to cross-check the
$\mathcal{L}_{\text{eff}}$ results and make sure that they were
robust.  Values measured for $\mathcal{L}_{\text{eff}}$ in both data
acquisition modes were consistent.

\item In contrast to this, the recent $\mathcal{L}_{\text{eff}}$
measurements described in Aprile {\it et al.} \cite{aprile} used a
trigger based only on scintillation light.  Detail is lacking on the
Monte Carlo used to generate the trigger efficiency correction and the
uncertainty in the inputs to the simulation, which would contribute to
the uncertainty in $\mathcal{L}_{\text{eff}}$.  Uncertainties
apparently not taken into account in that measurement include
uncertainty in nuclear recoil energy resolution, and systematic
uncertainty in subtracting the large multiple scattering background, 
which is dominated by events with small S1 signal. 
We note that because of the bulkiness of the liquid xenon cell used in
\cite{aprile} and the significant amount of inactive liquid xenon and
PTFE surrounding the active liquid xenon volume, multiple
scattering background is much more significant in the Columbia
measurement \cite{aprile} than in the Yale measurement \cite{manzur}. 
Overall, a number of contributions to systematic error are ignored in
the Aprile {\it et al.} \cite{aprile} analysis, and we feel that
that the claimed systematic errors are clearly underestimated.

\item  For nuclear recoils in LXe, the charge yield per keV is found to increase
as energy decreases. This very likely takes away signal that otherwise
could have gone toward production of scintillation light. It is
wishful thinking to presume that the scintillation signal remains
constant while the charge yield increases. See the empirical model
described in Manzur {\it et al} \cite{manzur}.

\item We called attention to the two-body kinematic cutoff (generating
an E$_{max}$ in LXe of 39 keV) not to claim that this is the only
effect relevant to liquid xenon scintillation, but merely to point out
that below this energy, it starts to become kinematically unfavorable
for nuclear recoils to ionize xenon, which normally is the main
process that leads to scintillation.  In models of scintillator
response, the scintillation yield does not drop immediately to zero,
but begins to adiabatically decrease at this point.  This is mentioned
several times in our Comments, and is an effect consistent with data
from other scintillators as well \cite{v3}.  For LXe, in order for
$\mathcal{L}_{\text{eff}}$ to remain constant below 39 keV$_{r}$, new
unknown physical processes would have to be invoked to balance this
effect and the effect of charge straggling described in \cite{manzur}. 
Section 3 in \cite{reply} reads as an protracted discussion 
constructed
to argue against fundamental concepts, widely regarded as common to
all scintillators.  We also notice a preoccupation in \cite{reply} to
keep comparing the XENON100 claimed sensitivity with the presently
favored DAMA region: as we have stated in \cite{v3}, these same
fundamental concepts can amplify the existing uncertainty in the
NaI[Tl] quenching factors, affecting the position of the DAMA region
in WIMP phase space, and in some plausible scenarios displacing it away from
XENON100 constraints.

\item Referring to the discussion around Sorensen {\it et al.}
\cite{prevnim}, the first author on that paper has
significantly improved the analysis of the XENON10 nuclear recoil 
calibration data \cite{peter}.  The new analysis
shows a $\mathcal{L}_{\text{eff}}$ that decreases at lower energies. 
According to Sorensen, the lowest energy points (denoted by open
circles in \cite{peter}) are not a claim of rising
$\mathcal{L}_{\text{eff}}$ as stated by XENON100, but instead an
illustration of how $\mathcal{L}_{\text{eff}}$ can be mistakenly found
to rise if threshold effects are not properly taken into account. 

\end{enumerate}

At this point we believe that a reader inclined to follow the fine
details of this discussion has been provided with enough information
to develop an opinion on the uncertainties affecting XENON100
results.  However, one important argument remains to be made: it seems
discernible to the trained eye that the acceptance showed in Fig.\ 3
in \cite{xenon100} cannot explain on its own the rapidly decreasing
sensitivity to AmBe neutron recoils noticeable below $\sim$6 keV$_{r}$
in Fig.\ 2 in \cite{xenon100}, when taking into account that the
expected trend in such a calibration is a rapid rise in counting rate
at low recoil energy (see for instance Fig.\ 2 in \cite{v3}, this is
an expected behavior common to fast neutron irradiations of targets
comprising heavy nuclei).  This can be interpreted as an indication of
a decreasing $\mathcal{L}_{\text{eff}}$ with decreasing energy in
XENON100, similar to that recently found for XENON10 \cite{peter}.

The present reluctance by the XENON100 collaboration to release a
comparison between expected and measured response to neutron-induced
nuclear recoils in their apparatus makes it very hard to quantify such
impressions.  Their attitude is isolated: it is customary in
experimental searches for WIMPs (e.g. CDMS \cite{cdms}, COUPP
\cite{coupp}, ZEPLIN \cite{zeplin}, XENON10 \cite{xenon10}, etc.) to
calibrate the detectors with fast neutrons, limiting the energy region
usable for the search to that for which an understanding of the
response exists, or alternatively folding into the analysis a
sensitivity penalty from any existing disagreements.  These can have a
fundamental origin as in the case of $\mathcal{L}_{\text{eff}}$, or an
instrumental explanation as in the case of the recoil signal
acceptance.  Hence the importance of performing these calibrations
{\it in situ}, using the same device dedicated to the WIMP search. 
We invite the XENON100 collaboration to revert to the mainstream by
adopting these conservative practices.  After extensive consultations, we
believe to be speaking for the majority of the dark matter
experimental community when making this request.  This will result in
a recovered credibility for present and future XENON100 claims.  The
alternative would involve a sort of magical thinking: to expect sensitivity
to WIMP-induced nuclear recoils, when the response to neutron recoils
of the same energy is weak or absent.


\begin{thebibliography}{10}

\bibitem{reply} XENON100 collaboration, {\tt arXiv:1005.2615}.
\bibitem{v3} J.I. Collar and D.N. McKinsey, {\tt arXiv:1005.0838}.
\bibitem{xenon100} E. Aprile {\it et al.}, {\tt arXiv:1005.0380v1}.
\bibitem{manzur}A. Manzur {\it et al.}, Phys. Rev. {\bf C81} 
(2010) 025808, {\tt arXiv:0909.1063}.
\bibitem{aprile} E. Aprile {\it et al.}, Phys. Rev. {\bf C79} (2009) 
045807.
\bibitem{prevnim} P. Sorensen {\it et al.}, Nucl. Instr. Meth. {\bf 
A601} (2009) 339, {\tt arXiv:0807.0459}.
\bibitem{peter} P. Sorensen (XENON10 collaboration), presented at the 
2010 Light Dark matter Workshop, UC Davis, available from
http://particle.physics.ucdavis.edu/seminars/  \\ 
data/media/2010/apr/sorensen.pdf
\bibitem{cdms} D.S. Akerib {\it et al.}, Phys. Rev. {\bf D72} (2005) 
052009, {\tt astro-ph/0507190v1}; Z. Ahmed {\it et al.}, Science {\bf 327}, (2010) 1619, 
{\tt arXiv:0912.3592}.
\bibitem{coupp} E. Behnke {\it et al.}, Science {\bf 319} (2008) 933, 
{\tt arXiv:0804.2886}.
\bibitem{zeplin} V.N. Lebedenko {\it et al.}, Phys. Rev. {\bf D80} (2009) 
052010,{\tt arXiv:0812.1150}.
\bibitem{xenon10} J. Angle {\it et al.}, Phys. Rev. Lett. {\bf 100} 
(2008) 021303, {\tt arXiv:0706.0039}.

\end{thebibliography}
\end{document}